\documentclass[twocolumn,prl,showpacs,preprintnumbers,amsmath,amssymb]{revtex4}
\usepackage{graphicx}

\begin{document}

\title{Bound on Lorentz- and CPT-Violating Boost Effects
for the Neutron}

\author{F.\ Can\`e, D.\ Bear, D.F.\ Phillips, M.S.\ Rosen, C.L.\ Smallwood,
R.E.\ Stoner and R.L.\ Walsworth}
\affiliation{
Harvard-Smithsonian Center for Astrophysics, Cambridge,
Massachusetts 02138}

\author{V.\ Alan Kosteleck\'y}
\affiliation{
Physics Department, Indiana University, Bloomington, Indiana 47405}

\date{Accepted for publication in Physical Review Letters on 10/25/2004.}

\begin{abstract}
  A search for an annual variation of a daily sidereal modulation of
  the frequency difference between co-located ${}^{129}$Xe and
  ${}^{3}$He Zeeman masers sets a stringent limit on boost-dependent
  Lorentz and CPT violation involving the neutron, consistent with no
  effect at the level of 150 nHz.  In the framework of the general
  Standard-Model Extension, the present result provides the
  first clean test for the fermion sector of the symmetry of spacetime
  under boost transformations at a level of $10^{-27}$~GeV.
\end{abstract}

\maketitle

Invariance of the laws of nature under transformations between
inertial reference frames that differ in relative velocity (``boosts'')
is an essential feature of Lorentz symmetry, along with
invariance under rotations.
Experimental tests of boost invariance, such as the Kennedy-Thorndike
experiment  \cite{ken}, have been performed for many years with
increasing precision  \cite{moreken}. These experiments typically
search for a variation of the velocity of light with
the laboratory velocity and test boost invariance in the context of
classical electrodynamics.
However, the fundamental role of Lorentz symmetry
in physics makes it desirable to test boost invariance for other systems,
such as massive particles with spin.

In this Letter, we present a high-sensitivity experimental test
of boost invariance for the neutron, which we interpret in the context
of
the Standard-Model Extension (SME) \cite{ck}, a general theoretical
framework that allows a comprehensive and systematic study of the
implications of Lorentz symmetry violation at observable energies.
The SME provides a widely-accepted formalism for the interpretation
and comparison of experimental
searches for violations of Lorentz symmetry and associated violations
of CPT symmetry (the product of Charge conjugation, Parity inversion,
and Time reversal). The SME
has been applied to many systems, including mesons
\cite{hadronexpt}, photons \cite{moreken,photonexpt}, and leptons
\cite{eexpt1,eexpt2,muons}, as well as the neutron
\cite{bear00,ccexpt} and proton \cite{phi}.
Observable Lorentz violation could be a remnant of Planck-scale
physics.  One attractive origin is spontaneous Lorentz breaking in a
fundamental theory \cite{kps}, but other sources are possible
\cite{overview}.

Our experiment consists of long-term monitoring of the frequency of
co-located  ${}^3$He and ${}^{129}$Xe Zeeman masers as the Earth
rotates and revolves around the Sun. We search for a specific
signature of a violation of boost invariance: an annual variation of
the nuclear Zeeman splitting, modulated at the frequency of the Earth's
daily sidereal rotation.  Such an effect could arise from couplings of the
${}^3$He and ${}^{129}$Xe nuclear spins
(each largely determined by a valence neutron)
to background tensor fields, including a dependence of the Zeeman
frequencies on the instantaneous velocity (magnitude and direction)
of the laboratory. 
The appeal of the noble-gas maser experiment is the excellent absolute
frequency stability \cite{stoner96,bear98,bear00}, and thus the
sensitivity to small, slow variations in the magnitude of
Lorentz-violating spin couplings.

Using the two-species noble-gas
maser, we recently constrained the possible
rotation-symmetry-violating couplings of the neutron spin with respect
to an inertial reference frame based on the Earth \cite{bear00}. Here, we
choose a Sun-based inertial reference frame, which allows us to study
cleanly --- for the first time in the fermion sector --- the symmetry
of spacetime with respect to {\it boost} transformations.
(The pioneering work of Berglund {\it et al.} \cite{cc6} does not
distinguish between the boost and rotation effects to which it is
sensitive.)
Our experiment's rest frame moves with the Earth around the Sun at a
velocity of magnitude $v_{\oplus}/c=\beta_{\oplus} \simeq 9.9\times
10^{-5}$, and the Lorentz transformation that describes the change of
coordinates from the laboratory frame to the Sun-based frame includes
both a rotation, $\mathbf{R}$, and a boost along the velocity
$\vec{\beta}$.

The most general, coordinate-independent Hamiltonian, $H$, containing
the Zeeman effect (from an applied magnetic field $\vec{B}$) and
Lorentz-symmetry-violating couplings of the noble gas nuclear spins,
$\vec{I}$, including leading terms
to first order
in $\vec{\beta}$, takes the simple form
\begin{equation} 
H = \vec{I} \cdot \left(
\gamma \vec{B} + 
\mathbf{R}(t) \vec{\lambda_{\odot}} +
\vec{\beta}(t)  \mathbf{R}(t) \mathbf{\Lambda_{\odot}} 
\right).
\label{GEN}
\end{equation}
Here, the vectors $\vec{I}$ and $\vec{B}$ are expressed in the lab
frame; 
whereas the explicit Lorentz-symmetry-violating vector
$\vec{\lambda_{\odot}}$ and $3\times3$ matrix
$\mathbf{\Lambda_{\odot}}$ have elements that are combinations of SME
coefficients, which may be determined in terms of fundamental
Lorentz-violating interactions \cite{kos1,blu2}, and are assumed
constant in the Sun frame.
The second term of Eq.~(\ref{GEN}) leads to a rotation-dependent
modulation of the maser frequency.
The third term contains cross-couplings in which the rotation induces
daily sidereal modulations of the maser frequencies, while the boost
transformation induces a sinusoidal variation of the daily modulation
amplitude over the course of the sidereal year \cite{sid} as the
direction of the velocity of the Earth varies with respect to the Sun.
Terms from higher rank tensors (such as a yearly modulation in the
maser frequency -- for which our maser does not have the stability to
set strong limits) have been neglected.

We refer the reader to previous
publications~\cite{stoner96,bear98,bear00} for details on the design
and operation of our two-species noble-gas Zeeman maser.  Here, we
provide a brief review. Co-located ensembles of ${}^{129}$Xe and
${}^{3}$He atoms at pressures of hundreds of mbar are held in a
double-chamber glass cell placed in a homogeneous magnetic field of
$\sim 1.5$ G.  Both species have spin-1/2 nuclei and the same sign
nuclear magnetic dipole moment, but no higher-order electric or
magnetic nuclear multipole moments. In one chamber of the glass cell,
the noble gas atoms are nuclear-spin-polarized by spin-exchange
collisions with optically-pumped Rb vapor \cite{optPump}. The noble
gas atoms diffuse into the second chamber, which is surrounded by an
inductive circuit resonant both at the ${}^{3}$He and ${}^{129}$Xe
Zeeman frequencies (4.9 kHz and 1.7 kHz, respectively).  For a
sufficiently high flux of population-inverted nuclear magnetization,
active maser oscillation of both species can be maintained
indefinitely.

Due to the generally weak interactions of noble gas atoms with the
walls and during atomic collisions, the ${}^{3}$He and ${}^{129}$Xe
ensembles can have long Zeeman coherence ($T_{2}$) times $\sim$
hundreds of seconds. It is thus possible to achieve excellent absolute
frequency stability with one of the noble-gas masers by using the
second maser as a co-magnetometer. For example, Zeeman frequency
measurements with sensitivity of $\sim 100$~nHz are possible with
averaging intervals of about an hour \cite{bear98}. This two-species
noble gas maser can also serve as a sensitive NMR
gyroscope~\cite{navigation}: the above quoted frequency stability
implies a rotation sensitivity of $0.13$ degree/hour.

For the boost-symmetry test, we choose a set of laboratory coordinates
$(t,x,y,z)$ such that the $\hat{x}$ axis points south, the $\hat{y}$
axis points east, and the $\hat{z}$ axis points vertically upwards in
the laboratory \cite{chg}.  With the reasonable approximation that the
orbit of the Earth is circular, the rotation, $\mathbf{R}$, from the
Sun-centered celestial equatorial frame to the standard laboratory
frame is given by
\begin{equation} \label{rot} R^{jJ}=\left( \begin{array}{c c c}
\cos\chi~\cos\omega_{\oplus}T_{\oplus} & \cos\chi~ \sin
\omega_{\oplus}T_{\oplus} & -\sin \chi \\
-\sin \omega_{\oplus}T_{\oplus} &  \cos\omega_{\oplus}T_{\oplus} & 0 \\
\sin \chi~ \cos \omega_{\oplus}T_{\oplus} &  \sin \chi
~\sin\omega_{\oplus}T_{\oplus} &
\cos \chi \\ \end{array}
\right). \end{equation}
In this equation, $j=x,y,z$ denotes the spatial index in the laboratory
frame, while $J=X,Y,Z$ denotes the spatial index in the Sun-centered
frame using celestial equatorial coordinates. The Earth's sidereal angular
rotation frequency is
$\omega_{\oplus} \simeq 2 \pi /$(23 h 56 min),
and $\chi\simeq 47.6^{\circ}$ is the colatitude of the
laboratory, located in Cambridge, Massachusetts. The time $T_{\oplus}$ is
measured in the Sun-centered frame from the beginning of the sidereal day,
which begins when the $\hat{y}$ and $\hat{Y}$ axes align.

The velocity 3-vector of the laboratory in the Sun-centered frame is
\begin{equation} 
  \label{vel} 
  \vec{\beta}=\beta_{\oplus} \left( 
      \sin \Omega_{\oplus}T,
      -\cos \eta ~\cos \Omega_{\oplus}T,
      -\sin \eta ~\cos \Omega_{\oplus}T  
  \right). 
\end{equation}
Here, $\Omega_{\oplus}$ is the angular frequency of the Earth's orbital
motion. The time $T$ is measured by a clock at rest at the origin,
with $T=0$ taken at 2:35
a.m. (U.S. Eastern Standard Time), March 20, 2000 \cite{equ}.
The angle between the $XY$
celestial equatorial plane and the Earth's orbital plane
is $\eta=23.4^{\circ}$.
We have ignored the laboratory's velocity due to the rotation of the
Earth, whose magnitude, $\beta_{L}=r_{\oplus} \omega_{\oplus} \sin\chi
/c \simeq 1.1 \times 10^{-6}$ (where $r_{\oplus}$ is the radius of the
Earth), is two orders of magnitude smaller than the orbital velocity.

We assume that 
the Lorentz-violating coefficients of $\vec{\lambda_{\odot}}$ and
$\mathbf{\Lambda_{\odot}}$ 
are static and spatially uniform
in the Sun frame, at least over the course of a solar year.
The corresponding coefficients
in the laboratory frame thus acquire a time dependence
due to both the Earth's rotation and its revolution around the Sun.
We also assume observer Lorentz covariance; hence
direct Lorentz transformations yield the coefficients in the
laboratory frame.

In the boost-symmetry test, we used the ${}^{129}$Xe maser as a
co-magnetometer to stabilize the magnetic field, which was oriented
along the $y$ axis (i.e., east-west). Thus the leading
Lorentz-violating frequency
variation of the free-running ${}^{3}$He maser was given
by:
\begin{equation} \delta\nu_{\mathrm{He}} =
\delta\nu_{X}~\sin~\omega_{\oplus}T_{\oplus}+
\delta\nu_{Y}~\cos~\omega_{\oplus}T_{\oplus},
\label{fiteq}
\end{equation}
where
\begin{eqnarray}
\delta\nu_{X}=k~(\lambda_{s}+\beta_{\oplus}(\Lambda_{ss}\sin{\Omega_{\oplus}T}+
\Lambda_{sc}\cos{\Omega_{\oplus}T})),
\label{nux} \\
\delta\nu_{Y}=k~(\lambda_{c}+\beta_{\oplus}(\Lambda_{cs}\sin{\Omega_{\oplus}T}+
\Lambda_{cc}\cos{\Omega_{\oplus}T})).
\nonumber
\end{eqnarray}
Here $\lambda_{c}$, $\lambda_{s}$, $\Lambda_{ss}$,
$\Lambda_{sc},\ldots$ are combinations of Sun-frame
Lorentz-violating coefficients of $\lambda_{\odot}$ and
$\Lambda_{\odot}$;
and $k=-8.46 \times 10^{32}$~nHz/GeV \cite{bear00}.

We note that Eqs.~(\ref{fiteq}) and (\ref{nux}) cleanly distinguish
the effects of rotation alone (terms proportional to $\lambda_{c}$ and
$\lambda_{s}$) from the effects of boosts due to the Earth's motion
(terms proportional to $\Lambda_{cc}$, $\Lambda_{cs}$, $\Lambda_{sc}$,
$\Lambda_{ss}$). In addition, these equations indicate, that the
sensitivity of our experiment to violations of boost-symmetry is
reduced by a factor of $\beta_{\oplus} \simeq 10^{-4}$ with respect to
the sensitivity to rotation-symmetry violation. However, for various
models of Lorentz violation that are isotropic in the frame of the
cosmic microwave background \cite{natural}, our experiment has greater
sensitivity to boost-symmetry violation than to rotation-symmetry
violation.

\begin{table}
\begin{center}
\begin{ruledtabular}
\begin{tabular}{ccc}
$T_{j}$ & $\delta \nu_{X}$~(nHz) & $\delta
\nu_{Y}$~(nHz) \\
\hline
4/15/99 & $156 \pm 90$ & $37 \pm$ 90 \\
9/15/99 & $-100 \pm 112$ & $-162 \pm 148$  \\
3/10/00 &  $42 \pm 86$ & $25 \pm 76$ \\
4/22/00 & $125 \pm 80$ & $-25 \pm 99$
\end{tabular}
\end{ruledtabular}
\caption{Mean and standard error of the two quadratures $\delta\nu_{X}$ and
$\delta\nu_{Y}$ of the sidereal day ${}^{3}$He-maser frequency modulations,
for each of four runs.  $T_{j}$ indicates the mean date of the $j$-th run.}
\label{t.data}
\end{center}
\end{table}

\begin{figure}
\includegraphics[width=0.73\hsize]{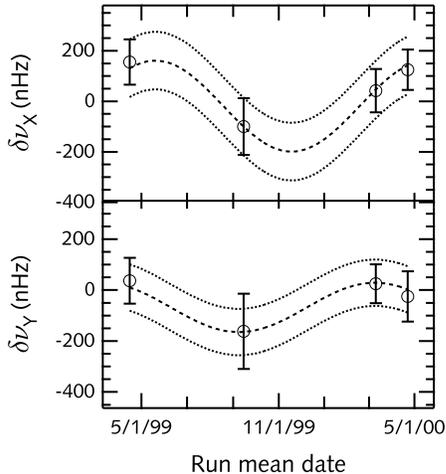}
\smallskip
\caption{Time course of the mean values of $\delta\nu_{X}$ and
  $\delta\nu_{Y}$.  For each plot the dashed line is the best fit
  obtained from Eq.~(\ref{nux}), using the fit parameters
  $\lambda_{c}$, $\lambda_{s}$, $\Lambda_{cc}$, $\Lambda_{cs}$,
  $\Lambda_{sc}$, $\Lambda_{ss}$ (see also the fit results in Table
  \ref{SMEParameters}). Dotted lines indicate the $1$$\sigma$
  confidence bands for the fit model.}
\label{fig1}
\end{figure}

As discussed in \cite{bear00}, we acquired noble-gas maser data in four
different runs spread over about 13 months (see Fig.~\ref{fig1}). Each
run lasted about 20 days, and we reversed the direction of the
magnetic field after the first $\sim~10$ days in each run to help
distinguish possible Lorentz-violating effects from diurnal systematic
variations. We fit this data to Eq.~(\ref{fiteq}). Table \ref{t.data}
lists, for each run, the mean values we determined for $\delta\nu_{X}$
and $\delta\nu_{Y}$, the amplitudes of sidereal-day modulations of the
${}^{3}$He-maser frequency due to Lorentz-violating coefficients in
the $\hat{X}$ and $\hat{Y}$ directions (Sun-centered frame). For each
run, $\delta\nu_{X}$ and $\delta\nu_{Y}$ correspond to a very good
approximation to a single high-precision measurement of the $X$ and
$Y$ components of $\delta\nu_{\mathrm{He}}$ performed at the mean time $T_{j}$.

Next, we fit the values of $\delta\nu_{X}$, $\delta\nu_{Y}$, and $T_{j}$ in
Table \ref{t.data} to Eq.~(\ref{nux}), thus obtaining the fit
parameters reported in Table \ref{SMEParameters} and shown
graphically in Fig.~(\ref{fig1}).
We treated all fit parameters as 
independent and we extracted energy bounds for 
Lorentz-violating
coefficients disregarding the possibility of accidental mutual
cancellations. This analysis yielded no significant violation of boost
invariance, with a limit of about $150$ nHz on the magnitude of an
annual modulation of the daily sidereal variation in the
${}^{3}$He-maser frequency.

To confirm that our result is consistent with the null hypothesis
(i.e., no Lorentz-violating effect), we performed two checks.  First,
we generated $10,000$ faux ${}^{3}$He-maser data sets including
sidereal-day frequency variations drawn from a normal distribution of
zero mean, but with standard errors for $\delta\nu_{X}$ and
$\delta\nu_{Y}$ at each time $T_{j}$ equal to the corresponding values
found in the experiment.  For each faux data set, we calculated the
$\chi^{2}$ of the fit to Eq.~(\ref{nux}), and found that the value
$\chi^{2}=0.30$ from the real experimental data is highly probable for
a system in which there is no daily sidereal modulation of the
${}^{3}$He maser frequency at the experiment's level of sensitivity.
In the second check, we performed a series of F tests to estimate the
probabilities that the values of the fit parameters, determined from
the maser data, arise entirely from statistical fluctuations. For all
fit parameters, the F tests yielded probabilities greater than $30\%$;
whereas it is customary to consider that a fit parameter is
significantly different from zero only if the F test probability is
smaller than $5\%$ or $1\%$.

We also performed a series of checks for systematic effects, including
sidereal-day and -year variations in maser temperature and signal amplitude
(e.g., driven by variations in the optical-pumping laser). Temperature
fluctuations in the ${}^{3}$He and ${}^{129}$Xe
detection circuit
can induce small maser-frequency shifts. Accurate temperature
monitoring over the
course of the 13-month experiment showed a maximum $1.6$ mK sidereal-day
variation of maser temperature, corresponding to a maximum sidereal-day
${}^{3}$He-maser frequency modulation of about $4$ nHz, which is an
order of magnitude smaller than our statistical sensitivity. A careful
analysis of the maser amplitude showed a lack of phase coherence in
sidereal-day modulations over the 13-month data set, and hence an insignificant
systematic sidereal-year variation in the ${}^{3}$He-maser frequency.

\begin{table*}
\begin{center}
\begin{ruledtabular}
\begin{tabular}{ccc}
Measurable combinations of SME coefficients& Fit parameters & Fit
results (GeV)\\
\hline
$\tilde{b}_{Y}-0.0034\tilde{d}_{Y}+0.0034\tilde{g}_{DY}$ & $\lambda_{c}$ &
$(8.0\pm 9.5)~10^{-32}$ \\
$-\tilde{b}_{X}+0.0034\tilde{d}_{X}-0.0034\tilde{g}_{DX}$ & $\lambda_{s}$ &
$(2.2\pm 7.9)~10^{-32}$\\
$-\cos \eta [(\frac{1}{2} \tilde{b}_{T}+\frac{1}{2}
\tilde{d}_{-}-\tilde{g}_{c}-\frac{1}{2} \tilde{g}_{T})+(\tilde{g}_{T}
-2\tilde{d}_{+}+\frac{1}{2}
\tilde{d}_{Q})]  +\sin \eta (\tilde{d}_{YZ}-\tilde{H}_{XT})$ &
$\Lambda_{cc}$ &
$(-1.1\pm 1.0)~10^{-27}$\\
$-\tilde{H}_{ZT}$ & $\Lambda_{cs}$ & $(0.2\pm 1.8)~10^{-27}$ \\
$[(\frac{1}{2} \tilde{b}_{T}+\frac{1}{2}
\tilde{d}_{-}-\tilde{g}_{c}-\frac{1}{2} \tilde{g}_{T})-(\tilde{g}_{T}
-2\tilde{d}_{+}+\frac{1}{2}
\tilde{d}_{Q})]$ & $\Lambda_{ss}$ & $(-1.8\pm 1.9)~10^{-27}$\\
         $\cos \eta(\tilde{H}_{ZT}-\tilde{d}_{XY})-\sin \eta
\tilde{H}_{YT}$ &$\Lambda_{sc}$ & $(-1.1\pm 0.8)~10^{-27}$\\
\end{tabular}
\end{ruledtabular}
\caption{
  Limits from the present work on Lorentz violation of
  the neutron, expressed in terms of (i) the fit parameters of
  Eqs.~(\protect\ref{fiteq}) and (\protect\ref{nux}), \emph{i.e.},
  coefficients for the general Lorentz-symmetry-violating vector
  $\vec{\lambda_{\odot}}$ and $\mathbf{\Lambda_{\odot}}$ (both in
  the Sun-frame); and (ii) combinations of Sun-frame SME coefficients
  for Lorentz and CPT violation (defined in Appendix B of
  Ref.~\protect\cite{blu2}). Bounds on rotation-symmetry violation are
  set by the limits on $\lambda_c$ and $\lambda_s$, whereas bounds on
  boost-symmetry violation are determined from $\Lambda_{cc}$,
  $\Lambda_{cs}$, $\Lambda_{ss}$, and $\Lambda_{sc}$.
} 
\label{SMEParameters}
\end{center}
\end{table*}

To interpret this test of boost invariance,
we follow the conventions of Ref.~\cite{kos2}, Appendix C, 
which allows us to
relate
the maser frequencies to the various SME coefficients for Lorentz and
CPT violation.
In 
particular, 
the neutron --- and hence the frequency of each noble-gas maser --- is
sensitive to Lorentz and CPT violation controlled by the SME
coefficients $b_{\Lambda}$, $d_{\Lambda \Sigma}$, $H_{\Lambda
  \Sigma}$, and $g_{\Lambda \Sigma \Gamma}$ \cite{kos1}.
Table \ref{SMEParameters} shows the corresponding bounds provided by
our experiment to combinations of Sun-frame SME coefficients,
including the clean limit of $\sim10^{-27}$ GeV on boost-violation.

In conclusion, we used co-located ${}^{3}$He and ${}^{129}$Xe Zeeman
masers to perform a high-sensitivity search for a violation of boost
invariance of the neutron. We found no significant sidereal annual
variation in the free-running ${}^{3}$He-maser frequency at a level of
approximately $150$ nHz. 
This result provides the first clean test of boost
symmetry for a fermion; and 
in the context of the general Standard-Model Extension,
places a bound of about $10^{-27}$ GeV on
11 previously unexplored coefficients among the 44 coefficients
describing possible leading-order Lorentz- and CPT-violating couplings
of the neutron.  Significant improvements may be possible with a
${}^{21}$Ne/${}^{3}$He Zeeman maser \cite{sto2}, with masers located
on a rotating table \cite{rot}, or with space-based clocks \cite{blu}.

We acknowledge support from NASA grants NAG8-1784, NAG3-2878,
NAG3-2194, and NAG8-1770, DOE grant DE-FG02-91ER40661, and the
Smithsonian Institution.


\begin{thebibliography}{38}
\bibitem{chg}
The coordinate system adopted here
differs from that used in our earlier work, Ref.\ \cite{bear00}.
\bibitem{equ} At this instant Earth crossed the equatorial plane
during the vernal equinox. $T$ and  $T_{\oplus}$ are equivalent, modulo
the sidereal day period.
\bibitem{natural} An example of such a model expressed in terms of
  Standard Model Extension coefficients consists of two background
  tensor fields ($b_{\Lambda}$ and $d_{\Lambda \Sigma}$) with the
  following simple relationships: $d_{XX}=d_{YY}=d_{ZZ},
  d_{TT}=3d_{XX}, b_{T}=4md_{XX}$, where $m$ is the mass of the
  neutron and all other terms are zero.  For this model, our
  experiment has an order of magnitude greater sensitivity to
  boost-symmetry violation than to rotation-symmetry violation,
  despite the factor of $\beta_{\oplus} \simeq 10^{-4}$.
\bibitem{rot}
Successful tests of Lorentz symmetry have been performed using
rotating tables.
See, e.g.,
Ref.\ \cite{eexpt2}.
\end{thebibliography}
\end{document}